# Existence and Stability of Equilibrium of DC Microgrid with Constant Power Loads

Zhangjie Liu, Mei Su, Yao Sun, Wenbin Yuan and Hua Han

*Abstract*— Constant power loads (CPLs) are often the cause of instability and no equilibrium of DC microgrids. In this study, we analyze the existence and stability of equilibrium of DC microgirds with CPLs and the sufficient conditions for them are provided. To derive the existence of system equilibrium, we transform the problem of quadratic equation solvability into the existence of a fixed point for an increasing fractional mapping. Then, the sufficient condition based on the Tarski fixed-point theorem is derived. It is less conservative comparing with the existing results. Moreover, we adopt the small-signal model to predict the system qualitative behavior around equilibrium. The stability conditions are determined by analyzing quadratic eigenvalue. Overall, the obtained conditions provide the references for building reliable DC microgrids. The simulation results verify the correctness of the proposed conditions.

*Index Terms*-- DC microgrid, solvability, nonlinear equations, fixed-point theorem, quadratic eigenvalue problem, stability, constant power load.

## I. Introduction

RECENTLY, DC microgrids have attracted much attention given three main advantages over their AC counterparts, namely high efficiency, simple control and robustness. They are increasingly used in applications such as aircraft, spacecraft and electric vehicle [1]–[3]. In DC microgrid, most loads are connected to the DC-bus through power electronic interface circuits which make the loads behave constant power loads (CPLs) [4].

Nevertheless, the negative impedance of CPLs often result in instability. Thus, stability has been widely investigated in DC microgrids with CPLs, and several stabilization methods have been proposed [5]–[20]. The topologies of DC microgrids in these studies can be divided into two groups regarding the number of converters and loads: 1) $n$ converters and one CPL; 2) $n$ converters and $m$ CPLs.

Various studies focused on stability criteria and stabilization methods of DC microgrids composed of $n$ converters and one CPL [5]-[19]. In this topology, the DGs are connected to the CPLs through a DC bus, and when the DC-bus resistance is neglected, the loads can be modeled as a single CPL. Therefore, this system is equivalent to star-connection topology consisting of $n$ DGs and one CPL. The existence of equilibrium in this system is determined by a quadratic equation with one unknown, and hence the sufficient conditions can be easily obtained [12], [19]. Therefore, related studies mainly focus on overcoming the instability due to CPL. Some linear techniques are used to stabilize the system [5]–[8]. Based on the idea that increasing damping can mitigate oscillations, several stabilization methods such as passivity based control [5], active damping [6], and virtual impedance [7]–[8] have been proposed. Furthermore, several nonlinear methods such as phase-plane analysis [9], feedback linearization [10] and sliding-mode control [11] have been applied to overcome CPL instability. To estimate the region of attraction around known equilibria, Lyapunov-like functions have been proposed, including Lure Lyapunov function [12], Brayton-Moser's mixed potential [13]–[14], block diagonalized quadratic Lyapunov function [15] and Popov criterion [16]. In addition, the stability of DC microgrids under droop control was analyzed in [17]–[19]. A stability condition is derived based on the reduced-order model of DC microgrids [17]. Another stability condition is derived under the assumption that all the DG filter inductances have the same ratio, $R/L$ [18]. To obtain more accurate stability conditions, a high dimensional model is used in stability analysis [19].

Due to the transmission loss, the increase of CPL may result in the loss of system equilibrium (i.e., voltage collapse) [20] [22]. Hence, finding the condition for the existence of the equilibrium is a prerequisite. The condition for equilibrium existence can be easily obtained under the assumption that the DC-bus resistance can be neglected. However, in most cases, this assumption is not reasonable. In the practical mesh DC-microgrid consisting of $n$ converters and $m$ CPLs, the existence of equilibrium is determined by an $m$-dimensional quadratic equation with $m$ unknowns, which is a difficult problem. By using quadratic mapping, a necessary condition for the existence of equilibrium based on a linear matrix inequality (LMI) is obtained in [20], where the condition is sufficient if and only if the quadratic mapping is convex. Although a test to verify the convexity of quadratic mappings is given in [21], the mappings are usually nonconvex for the case $m > 2$. To determine the sufficient condition for the solvability of the $m$-dimensional quadratic equation, a contraction mapping is constructed ingeniously in [22]. However, the obtained condition is a little conservative because contraction mapping requires that the norm of the mapping Jacbian matrix is less than 1.

In this study, we investigate the following two questions.
1) Under what conditions should the system admit a constant steady state and how can we reduce the conservativeness?
2) How to design a stabilization method be designed to guarantee system stability?

The main contributions of this paper can be summarized as
- A control method based on virtual resistance and virtual inductance concept to overcome instability is proposed.
- The sufficient condition for the system to admit an equilibrium is obtained, and it is less conservative compared with [22].

- The equilibrium stability is analyzed and the analytical stability conditions are determined using eigenvalue analysis.

The rest of this paper is organized as follows: Section II introduces some preliminaries and notations. Section III describes the basic models and the proposed control scheme. The existence of equilibrium for DC microgrids is presented in Section IV. The stability analysis and the sufficient conditions are detailed in Section V. Simulation results are presented in section VI. Finally, we draw our conclusions in Section VII.

## II. PRELIMINARIES AND NOTATION

**Definition 1.** We denote $A > 0$ if matrix $A$ is positive definite. Matrix $A$ (or a vector) is called nonnegative (respectively positive, negative and nonpositive) if its entries are nonnegative (respectively positive, negative and nonpositive) and we denote $A \succ B, A \succeq B, A \prec B, A \preceq B$ if the entries of $A-B$ are all positive, nonnegative, negative and nonpositive, respectively. In addition, $1_n$ ($0_n$) is the vector of ones (zeros). For Hermitian matrix $A \in R^{m \times m}$, we denote its eigenvalue as $\lambda_1(A) \leq \cdots \leq \lambda_m(A)$.

**Definition 2.** Square matrix $A$ is a Z-matrix if all the off-diagonal elements are zero or negative, and it is also an M-matrix [23] if and only if one of these statements is true:
1) the eigenvalues of $A$ are in the right half-plane;
2) there exists a positive vector, $x$, such that $Ax \succ 0_n$.

**Definition 3.** Matrix $A \in R^{n \times n}$ is irreducible if there exists no permutation matrix $P$ such that $P^T A P$ can be represented as

$$P^T A P = \begin{bmatrix} A_{11} & A_{12} \\ O & A_{22} \end{bmatrix}$$

where $A_{11}$, and $A_{22}$ are square matrices, and $O$ is the zero matrix of proper dimension [24].

**Lemma 1.** If $A$ is an M-matrix, then
1) $A + D$ is an M-matrix for every nonnegative diagonal matrix $D$;
2) if $A$ is irreducible, $A^{-1}$ is positive [25].

**Lemma 2.** *Tarski fixed-point theorem* [26]. Given $D \subset R^{n \times n}$ convex, let $f : D \to D$ be a continuous function such that
1) $f(x)$ is strictly increasing, i.e., $\forall x_1, x_2 \in D$, if $x_1 \succ x_2$, $f(x_1) \succ f(x_2)$;
2) $\exists x_1, x_2 \in D$ such that $x_1 \prec f(x_1) \prec f(x_2) \prec x_2$,

there is a unique vector $x_1 \prec x^* \prec x_2$ such that $f(x^*) = x^*$.

**Lemma 3.** Let $A$ be a real positive matrix. Perron root $\chi$ and Perron vector $\eta$ satisfy $A\eta = \chi\eta$, where $\eta \succ 0$ and $\eta^T\eta = 1$. Moreover, $\chi$ is also the spectral radius of $A$, denoted $\rho(A)$. If $A \succ B \succ O$, then $\rho(A) > \rho(B)$ [24].

**Lemma 4.** Let $Q(\lambda) = \lambda^2 M + \lambda K + C$. Let $M$, $K$ and $C$ be positive definite Hermitian matrices. For a quadratic eigenvalue problem $|\lambda^2 M + \lambda K + C| = 0$, Re $(\lambda) <0$ [27].

## III. DC-MICROGRID MODEL AND CONTROL SCHEME

A general DC microgrid with $n$ converters (DGs) and $m$ CPLs is illustrated in Fig. 1 and consists of three main components: sources, loads and cables. In a low-voltage DC microgrid, the cable can be regarded as purely resistive and we assume the loads as CPLs. In addition, we consider the DC-microgrid topology as a graph with the sources and loads representing nodes, and the cables representing edges, respectively. Furthermore, we assume the graph as being strongly connected, i.e., every source has access to every load.

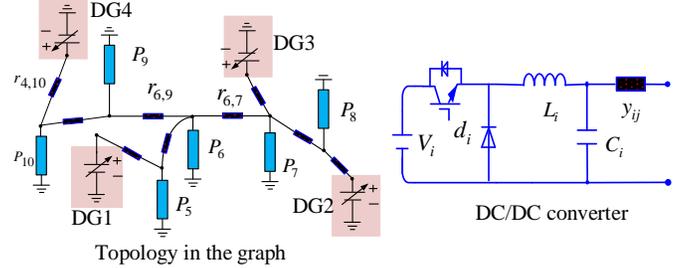

Fig.1. Diagram of the DC microgrid.

The dynamics of the $i^{th}$ converter can be described by

$$\begin{cases} L_i \dfrac{d\tilde{i}_{L_i}}{dt} = V_i d_i - u_i \\ C_i \dfrac{du_i}{dt} = \tilde{i}_{L_i} - i_i \end{cases} \quad (1)$$

where $V_i$, $d_i$, $\tilde{i}_{L_i}$, $u_i$ and $i_i$ are the input voltage, duty cycle, inductance current, output voltage and current of the converter, respectively.

To prevent the instability caused by CPLs in a DC microgrid, we propose a controller based on virtual resistance and virtual inductance whose duty cycle is given by

$$d_i = \frac{L_i}{V_i X_i}\left(u_{\text{ref}} - k_i \tilde{i}_{L_i} - u_i\right) + \frac{u_i}{V_i} \quad i \in \{1,2,\cdots,n\}, \quad (2)$$

where $X_i$, $k_i$, and $u_{\text{ref}}$ are the virtual inductance, virtual resistance, and reference voltage of the $i^{th}$ converter, respectively. Then, substituting (2) into (1) and representing the result in matrix form, we obtain

$$\begin{cases} X \dfrac{d\tilde{i}_L}{dt} = V^* - K\tilde{i}_L - u_S \\ C \dfrac{du_S}{dt} = \tilde{i}_L - i_S \end{cases} \quad (3)$$

where $\tilde{i}_L = \begin{bmatrix} \tilde{i}_{L_1} & \tilde{i}_{L_2} & \cdots & \tilde{i}_{L_n} \end{bmatrix}^T$, $u_S = [u_1\ u_2\ \ldots\ u_n]^T$, $i_S = [i_1\ i_2\ \ldots\ i_n]^T$, $V^* = u_{\text{ref}} 1_n$, $C = diag\{C_i\}$, and $X = \text{diag}\{X_i\}$, $i \in \{1, 2, \ldots, n\}$. In addition, the load voltages are denoted by $u_L = [u_{n+1}\ u_{n+2}\ \ldots\ u_{n+m}]^T$.

Next, applying the Kirchoff's and Ohm's laws, we have

$$i_i = \sum_{j=1}^{n+m} y_{ij}\left(u_i - u_j\right), \quad i \in \{1,2,\cdots,n+m\} \quad (4)$$

where $y_{ij}$ is the cable conductance, $y_{ij}=0$ if there is no cable connecting nodes $i$ and $j$, and $r_{ij} =1/y_{ij}$ represents the resistance. Writing (4) in block matrix form, we have

$$\begin{bmatrix} i_S \\ i_L \end{bmatrix} = \begin{bmatrix} Y_{SS} & Y_{SL} \\ Y_{LS} & Y_{LL} \end{bmatrix}\begin{bmatrix} u_S \\ u_L \end{bmatrix} = Y \begin{bmatrix} u_S \\ u_L \end{bmatrix}, \begin{cases} i_S = \begin{bmatrix} i_1 & i_2 & \cdots & i_n \end{bmatrix}^T \\ i_L = \begin{bmatrix} i_{n+1} & i_{n+2} & \cdots & i_{n+m} \end{bmatrix}^T \end{cases} \quad (5)$$

where $Y$ is the symmetric admittance matrix of the graph, $Y_{SS} \in R^{n \times n}$, $Y_{SL} \in R^{n \times m}$, $Y_{LS} \in R^{m \times n}$, and $Y_{LL} \in R^{m \times m}$ are the corresponding block matrices, $i_S$ and $i_L$ are the current vectors of the sources and loads, respectively. For a CPL, it yields

$$u_i i_i = -P_i, \ i \in \{n+1, n+2, \cdots, n+m\} \quad (6)$$

where $P_i$ is the power of load at node $i$, and the right-hand side of the equation is negative because the actual current direction and reference are opposite.

## IV. EXISTENCE OF EQUILIBRIUM OF DC MICROGRID

Besides instability, CPLs can cause inexistence of equilibrium in DC microgrids [19], [20]. For simplicity, all the loads can be equivalent to a common CPL under the assumption that the DC-bus resistance can be neglected [18]-[19]. Consequently, the existence of equilibrium is formulated by the solvability of a quadratic equation with one unknown, whose conditions are easily obtained. However, in most cases, the DC bus resistance cannot be neglected, and the equilibrium should be determined from quadratic equations with multiple unknowns. In this case, the topology of the solution set is complicated [20].

### A. Problem Formulation

According to (3), when the system achieves steady-state, the output voltage is given by $u_S = V^* - K i_S$, and substituting it into (5) yields

$$\begin{cases} i_S = Y_{SS} V^* - Y_{SS} K i_S + Y_{SL} u_L \\ i_L = Y_{LS} V^* - Y_{LS} K i_S + Y_{LL} u_L \end{cases}, \quad (7)$$

whose simplification yields to

$$\begin{cases} i_S = (Y_{SS}^{-1} + K)^{-1} V^* + (Y_{SS}^{-1} + K)^{-1} Y_{SS}^{-1} Y_{SL} u_L \\ i_L = (Y_{LS} - Y_{LS} K (Y_{SS}^{-1} + K)^{-1}) V^* + (Y_{LL} - Y_{LS} K (Y_{SS}^{-1} + K)^{-1} Y_{SS}^{-1} Y_{SL}) u_L \end{cases}. \quad (8)$$

Combining this result with (6), the system admits a constant steady state if and only if

$$\begin{cases} i_L = -\beta + Y_1 u_L \\ u_i i_i = -P_i \ i \in \{n+1, n+2, \cdots, n+m\} \end{cases} \quad (9)$$

is solvable where

$\beta = -u_{ref} Y_{LS} Y_{SS}^{-1} (Y_{SS}^{-1} + K)^{-1} 1_n, Y_1 = Y_{LL} - Y_{LS} K (Y_{SS}^{-1} + K)^{-1} Y_{SS}^{-1} Y_{SL}$.

Clearly, the system admits an equilibrium if and only if, for given values of $V^*$, $Y$, and $P$, the quadratic equations in (9) admit a real solution.

Let $U_L = diag\{u_L\}$, $P = [P_{m+1} \ P_{m+2} \ldots P_{m+n}]^T$, and $f = [f_1 \ f_2 \ldots f_m]^T = U_L(Y_{LS} V^* + Y_{LL} u_L)$. We define the multidimensional quadratic mapping, $f: R^m \to R^m$, of the form

$$f = [f_1(u_L) \ f_2(u_L) \cdots f_m(u_L)] = U_L(\beta + Y_1 u_L).$$

Then, $E = \{f(u_L): u_L \in R^m\}$ is the image of the space of variables $u_L$ under this map.

### B. Related Results

Some existed studies neglect the resistance of the common DC bus [18], [19]. Under this assumption, all CPLs in a DC microgrid are equivalent to a common CPL, as shown in Fig. 2, and (9) becomes

$$u_{eq}(\bar{Y}_{LS} V^* + \bar{Y}_{LL} u_{eq}) = -P_{eq} \quad (10)$$

where $P_{eq} = 1_m^T P$ is the equivalent load, $u_{eq} \in R$ is its voltage, $\bar{Y}_{LL} = 1_n^T (Y_{SS} + K^{-1})^{-1} 1_n$ and $\bar{Y}_{LS} = 1_n^T (Y_{SS} + K^{-1})^{-1}$ are the equivalent admittance matrices. This system admits an equilibrium if and only if

$$(\bar{Y}_{LS} V^*)^2 - 4\bar{Y}_{LL} P_{eq} \geq 0 \quad (11)$$

In fact, although (11) holds, system admits no constant steady-state equilibrium due to the resistance of the DC bus.

In [20], two results based on LMI are presented as:

**Proposition 1:** Assume there exists a diagonal matrix $H = diag\{h_i\}$ such that

$$\begin{cases} HY_1 + Y_1 H > 0 \\ 2\sum_{i=n+1}^{n+m} h_i P_i > (H\zeta)^T (HY_1 + Y_1 H)^{-1} H\zeta \end{cases} \quad (12)$$

Then, (9) has no real solution.

The necessary condition in Proposition 1 implies that if there exists a constant steady-state, LMI (12) has no solutions.

**Proposition 2:** If $E$ is convex, (9) have a real solution if and only if LMI (13) is unfeasible. Reference [21] provides a test to check convexity of $E$. However, for $m > 2$, set $E$ is usually nonconvex.

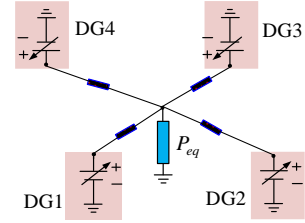

Fig.2. Equivalent topology of DC microgrid with $n$ DGs and one CPL.

In [22], a sufficient condition is obtained by using contraction mapping, i.e., the quadratic equations are solvable if

$$4\left\|(diag\{\zeta\} Y_1 diag\{\zeta\})^{-1} \Theta\right\|_\infty < 1 \quad (13)$$

where $\zeta = -Y_1^{-1} \beta$ and $\Theta = diag\{P_i\}$. The proof of (13) is detailed in [22], and here we provide an alternative proof along with two stronger conditions.

### C. Sufficient Conditions for Existence of Equilibrium

By simplifying (9), we have

$$U_L(\beta + Y_1 u_L) = -P \quad (14)$$

Multiplying by $Y_1^{-1} U_L^{-1}$, (14) becomes

$$u_L = F(u_L) = \zeta - Y_1^{-1} \Theta g(u_L), \ g(u_L) = [u_{n+1}^{-1} \ u_{n+2}^{-1} \cdots u_{n+m}^{-1}]^T \quad (15)$$

Then, the solvability of (14) is equivalent to the existence of fixed point in the fractional functions from (15). In addition, if 1) and 2) of Lemma 2 are satisfied, the system admits a constant steady state.

Now, we state the main results of this paper.

**Proposition 3:** The following two statements are true
1) $Y$ is irreducible;
2) $Y_1^{-1}$ is positive ($Y_1^{-1} \succ O$);
3) $\zeta = u_{ref} 1_m$.

**Proof.** First, we assume that $Y$ is reducible, and as $Y$ is symmetric, there exists a permutation matrix $E$ such that

$$E^T A E = \begin{bmatrix} Y_{11}' & O \\ O & Y_{22}' \end{bmatrix}. \quad (16)$$

Hence, the DC microgrid can be divided into two separate microgrids, which contradicts the assumption that all sources and loads are strongly connected, thus proving 1).

Next, $Y_1$ can be simplified as $Y_1 = Y_{LL} - Y_{LS}(Y_{SS} + K^{-1})^{-1} Y_{SL}$ and we define

$$\Gamma_1 \triangleq \begin{bmatrix} K^{-1} + Y_{SS} & Y_{SL} \\ Y_{LS} & Y_{LL} \end{bmatrix}. \tag{17}$$

Clearly, $Y_1$ is a Schur complement of $\Gamma_1$. According to 1) of Lemma 1, $\Gamma_1$ is a positive-definite M-matrix. Clearly, $\Gamma_1$ is irreducible as $Y$, and according to 2) of Lemma 1, $\Gamma_1^{-1}$ is strictly positive. Applying the formula for the inverse of a block matrix, we obtain

$$\Gamma_1^{-1} = \begin{bmatrix} (K^{-1} + Y_{SS} - Y_{SL}Y_{LL}^{-1}Y_{LS})^{-1} & -(K^{-1} + Y_{SS})^{-1} Y_{SL}Y_1^{-1} \\ -Y_{LL}^{-1}Y_{LS}(K^{-1} + Y_{SS} - Y_{SL}Y_{LL}^{-1}Y_{LS})^{-1} & Y_1^{-1} \end{bmatrix}.$$

Because $\Gamma_1^{-1}$ is strictly positive, $Y_1^{-1}$ is positive and irreducible, thus proving 2).

Finally, given that $Y$ is a Laplacian matrix, $Y1_{n+m} = 0_{n+m}$, i.e.,

$$\begin{cases} 1_n + Y_{SS}^{-1} Y_{SL} 1_m = 0_n \\ Y_{LL}^{-1} Y_{LS} 1_n + 1_m = 0_n \end{cases}. \tag{18}$$

Then, we have

$$(Y_{LS} - Y_{LS}K(Y_{SS}^{-1} + K)^{-1})1_n + (Y_{LL} - Y_{LS}K(Y_{SS}^{-1} + K)^{-1}Y_{SS}^{-1}Y_{SL})1_m \tag{19}$$
$$= (Y_{LS}1_n + Y_{LL}1_m) - Y_{LS}K(Y_{SS}^{-1} + K)^{-1}(1_n + Y_{SS}^{-1}Y_{SL}1_m) = 0_m$$

According to (18), we obtain $-u_{ref}^{-1}Y_1^{-1}\beta + 1_m = 0_m$, i.e., $\zeta = u_{ref} 1_m$, thus proving 3).

**Theorem 1.** A necessary condition for (15) to be solvable is

$$u_{ref} > 2\sqrt{\chi}, \tag{20}$$

where $\chi$ is the Perron root of $Y_1^{-1}\Theta$.

**Proof.** First, let $Y_1^{-1}\Theta = [a_1; a_2; \cdots; a_n]$ where $a_i$ is the $i^{th}$ row vector of $Y_1^{-1}\Theta$. We assume $\gamma \succ 0$ is a solution of (15) of the form

$$\gamma_i = u_{ref} - \sum_{j=1}^{m} a_{ij} \frac{1}{\gamma_j}, \ i \in \{1, 2, \cdots, m\}, \tag{21}$$

where $a_{ij}$ represents the vector entry. Thus, (21) can be expressed as

$$\left(\gamma_i - \frac{u_{ref}}{2}\right)^2 = \frac{u_{ref}^2}{4} - \gamma_i \sum_{j=1}^{m} a_{ij} \frac{1}{\gamma_j}, \ i \in \{1, 2, \cdots, m\}. \tag{22}$$

Then, the following can be obtained

$$\frac{u_{ref}^2}{4} \frac{1}{\gamma_i} - \sum_{j=1}^{m} a_{ij} \frac{1}{\gamma_j} > 0, \ i \in \{1, 2, \cdots, m\}, \tag{23}$$

whose matrix form is given by

$$(u_{ref}^2 I - 4Y_1^{-1}\Theta)\bar{\gamma} \succ 0_m, \bar{\gamma} = \begin{bmatrix} \gamma_1^{-1} & \gamma_2^{-1} & \cdots & \gamma_n^{-1} \end{bmatrix}^T \tag{24}$$

According to 2) of Proposition 3, $Y_1^{-1}\Theta$ is positive, i.e., $u_{ref}^2 I - 4Y_1^{-1}\Theta$ is a Z-matrix. In addition, (24) satisfies 2) of Definition 2, and hence $u_{ref}^2 I - 4Y_1^{-1}\Theta$ is an M-matrix. Therefore, according to 1) in Definition 2, (20) must be satisfied, completing the proof.

**Theorem 2.** There must exist a unique vector $u_L^*$ such that $u_L^* = F(u_L^*), h\xi \prec u_L^* \prec \zeta$ provided that

$$u_{ref}^2 > \max_{1 \le i, j \le m} \{f_{ij}(q)\} \tag{25}$$

where $q=[q_1 \ q_2 \ \ldots \ q_m]^T$ is an arbitrary positive vector and $h = \min\left\{\frac{q_i}{2}\left(u_{ref} + \sqrt{u_{ref}^2 - 4\frac{a_i q}{q_i}}\right)\right\}, \xi = \begin{bmatrix} q_1^{-1} & q_2^{-1} & \cdots & q_m^{-1} \end{bmatrix}^T$,

$$f_{ij}(q) = \begin{cases} 4\max\left\{\dfrac{a_i q}{q_i}, \dfrac{a_j q}{q_j}\right\} & \text{if } \dfrac{a_i q}{q_j} + \dfrac{a_j q}{q_i} \le 2\max\left\{\dfrac{a_i q}{q_i}, \dfrac{a_j q}{q_j}\right\} \\[2ex] \dfrac{\left(\dfrac{a_i q}{q_j} - \dfrac{a_j q}{q_i}\right)^2}{\dfrac{a_i q}{q_j} + \dfrac{a_j q}{q_i} - \dfrac{a_i q}{q_i} - \dfrac{a_j q}{q_j}} & \text{if } \dfrac{a_i q}{q_j} + \dfrac{a_j q}{q_i} > 2\max\left\{\dfrac{a_i q}{q_i}, \dfrac{a_j q}{q_j}\right\} \end{cases}$$

The proof is detailed in Appendix.

**Remark 1:** We transform the quadratic equation solvability into the existence of a fixed point in a nonlinear mapping. The key is to construct an appropriate mapping that satisfies the conditions of **Lemma 2**. In this process, the positivity of $Y_1^{-1}$ play a crucial role. However, for a passive-transmission network, its admittance matrix $Y$ is a symmetric positive semidefinite Z-matrix. Thus, $\Gamma_1$ and $Y_1$ must be irreducible M-matrices. Therefore, $Y_1^{-1}$ is positive, and function $F(x)$ is strictly increasing. This way, we can obtain the sufficient conditions for the system to admit an equilibrium by using the Tarski fixed-point theorem.

Furthermore, positive vector $q$ in (25) is arbitrary. Consequently, we can obtain the optimal $q$ that minimizes the right side of (25). Thus, the result is less conservative, and the optimal sufficient conditions can be formulated as

$$u_{ref}^2 > \min_q \left\{ \max_{1 \le i, j \le m} \{f_{ij}(q)\} \right\}, \tag{26}$$

Then, to obtain an explicit analytic condition, we take $q$ as several special vectors into (25).

**Corollary 1.** For given $Y$, $K$ and $P$, the system must admit a constant steady-state if the following holds

$$u_{ref} > \min\left\{ \frac{\bar{\eta} + \underline{\eta}}{\sqrt{\bar{\eta}\underline{\eta}}} \sqrt{\chi}, 2\sqrt{\|Y_1^{-1}\Theta\|_\infty} \right\}, \tag{27}$$

where $\eta$ is the Perron vector of $Y_1^{-1}\Theta$, $\bar{\eta} = \max\{\eta_i\}$ and $\underline{\eta} = \min\{\eta_i\}$.

**Proof.** Consider $q = 1_m$ in (25), and then

$$u_{ref}^2 > \max_{1 \le i, j \le m} \{f_{ij}(q)\} = \max_{1 \le i, j \le m} \{4\max\{a_i 1_m, a_j 1_m\}\} = 4\max_{1 \le i \le m}\{a_i 1_m\} = 4\|Y_1^{-1}\Theta\|_\infty. \tag{28}$$

Substituting $\zeta = u_{ref} 1_m$ into (13), it turns equivalent to (28). Thus, (13) is proved.

Let $q = \eta$, we have

$$\frac{a_i \eta}{q_j} + \frac{a_j \eta}{\eta_i} = \chi\left(\frac{\eta_i}{\eta_j} + \frac{\eta_j}{\eta_i}\right), \ 2\max\left\{\frac{a_i \eta}{\eta_i}, \frac{a_j \eta}{\eta_j}\right\} = 2\chi. \tag{29}$$

According to (29), $f_{ij}(\eta)$ becomes

$$f_{ij}(\eta) = \begin{cases} 4\max\left\{\dfrac{\chi\eta_i}{\eta_j}, \dfrac{\chi\eta_j}{\eta_i}\right\} = 4\chi & \text{if } \eta_i = \eta_j \\ \dfrac{\chi\left(\dfrac{\eta_i}{\eta_j} - \dfrac{\eta_j}{\eta_i}\right)^2}{\left(\sqrt{\dfrac{\eta_i}{\eta_j}} - \sqrt{\dfrac{\eta_j}{\eta_i}}\right)^2} = \chi\left(\sqrt{\dfrac{\eta_i}{\eta_j}} + \sqrt{\dfrac{\eta_j}{\eta_i}}\right)^2 & \text{if } \eta_i \neq \eta_j \end{cases}, \quad (30)$$

and hence the following is straightforward

$$\max\{f_{ij}(\eta)\} = \frac{(\bar{\eta} + \underline{\eta})^2}{\bar{\eta}\underline{\eta}} \chi. \quad (31)$$

Thus, (27) is obtained, completing the proof.

**Remark 2**: **Corollary 1** shows that for fixed load $P$, the system must admit a constant steady-state as long as $u_{\text{ref}}$ is large enough, as can be expected. Meanwhile, (26) and (27) provide numerical and analytical condition to guarantee the existence of the equilibrium, respectively. Moreover, both (26) and (27) are less conservative than the results from the recent works [22].

The relation among line resistances, virtual resistances, loads and DG output voltages that make (9) solvable is described by (26) and (27). This allows to determine design guideline to build reliable DC microgrid.

## V. STABILITY ANALYSIS OF DC MICROGRID

### A. Small-signal Model around Equilibrium

According to Theorem 2, if condition (26) holds, the system will admit a constant steady state, we denote it by $(\tilde{i}_L^*, u_S^*, i_S^*, i_L^*, u_L^*)$. Linearizing (6) around the equilibrium, the equivalent CPL resistances can be obtained from

$$\Delta i_i = \frac{1}{r_i}\Delta u_i, r_i = -P_i^{-1}(u_i^*)^2 \quad i \in \{n+1, n+2, \cdots, n+m\} \quad (32)$$

where $\Delta(\cdot)$ represents the small-signal variation around the equilibrium and $r_i$ is the negative CPL resistance. Substituting (32) in linearized (5), we have

$$\Delta i_L = \left(Y_{SS} - Y_{SL}\left(Y_{LL} + R_L^{-1}\right)^{-1} Y_{LS}\right)\Delta u_S, \quad (33)$$

where $R_L = \text{diag}\{r_i\}$, and combining the linearized (5) with (33), the equivalent linearized model of the system is given by

$$\begin{cases} X\dfrac{d\Delta\tilde{i}_L}{dt} = -K\Delta\tilde{i}_L - \Delta u_S \\ C\dfrac{d\Delta u_S}{dt} = \Delta\tilde{i}_L - Y_{eq}\Delta u_S \end{cases}, \quad (34)$$

where $Y_{eq} = Y_{SS} - Y_{SL}\left(Y_{LL} + R_L^{-1}\right)^{-1} Y_{LS}$. The system Jacobian matrix is given by

$$J_2 = \begin{bmatrix} -X^{-1}K & -X^{-1} \\ C^{-1} & -C^{-1}Y_{eq} \end{bmatrix}. \quad (35)$$

### B. Stability Conditions and Stabilization

According to the Hartman–Grobman theorem, the equilibrium is stable if and only if $J_2$ is Hurwitz. The characteristic polynomial of $J_2$ is obtained as

$$\left|\lambda I - \begin{bmatrix} -X^{-1}K & -X^{-1} \\ C^{-1} & -C^{-1}Y_{eq} \end{bmatrix}\right| = \left|\begin{matrix} \lambda I + X^{-1}K & X^{-1} \\ -C^{-1} & \lambda I + C^{-1}Y_{eq} \end{matrix}\right|$$

$$= \left|\lambda I + X^{-1}K\right|\left|\lambda I + C^{-1}Y_{eq} + C^{-1}\left(\lambda I + X^{-1}K\right)X^{-1}\right| \quad (36)$$

$$= \left|\left(\lambda I + X^{-1}K\right)\left(\lambda I + C^{-1}Y_{eq}\right) + I\right|$$

which results in

$$\left|\lambda^2 I + \lambda\left(X^{-1}K + C^{-1}Y_{eq}\right) + X^{-1}KC^{-1}Y_{eq} + X^{-1}C^{-1}\right| = 0. \quad (37)$$

To maintain symmetry in (37), we take $X = bK$, where $b$ is a proportionality coefficient. Then, the duty cycle is designed as

$$d_i = \frac{L_i}{bV_i k_i}\left(u_{\text{ref}} - k_i\tilde{i}_{L_i} - u_i\right) + \frac{u_i}{V_i} \quad i \in \{1, 2, \cdots, n\}, \quad (38)$$

and the system Jacobian matrix is

$$J_2 = \begin{bmatrix} -\dfrac{1}{b}I & -\dfrac{1}{b}K^{-1} \\ C^{-1} & -C^{-1}Y_{eq} \end{bmatrix}. \quad (39)$$

Then, simplifying (37), we have

$$\left|\lambda^2 I + \lambda\left(\frac{1}{b}I + C^{-1}Y_{eq}\right) + C^{-1}Y_{eq} + \frac{1}{b}K^{-1}C^{-1}\right| = 0 \quad (40)$$

**Theorem 3**. Matrix $J_2$ is Hurwitz if the following holds

$$\begin{cases} C + bY_{eq} > 0 \\ K^{-1} + bY_{eq} > 0 \end{cases} \quad (41)$$

**Proof.** Multiplying (37) by $C$, we have

$$\left|\lambda^2 C + \lambda\left(\frac{1}{b}C + Y_{eq}\right) + Y_{eq} + \frac{1}{b}K^{-1}\right| = 0 \quad (42)$$

Then, according to Lemma 3, (41) can be easily obtained. Combining with the existence condition of equilibrium point, the system admit a stable operation point if (26) and (41) hold.

**Proposition 4:** The minimal eigenvalue $\lambda_1(Y_{eq})$ of $Y_{eq}$ is negative.

**Proof.** Clearly, $Y_{eq}$ is the Schur complement of $\Gamma_2$ which is defined as

$$\Gamma_2 = \begin{bmatrix} Y_{SS} & Y_{SL} \\ Y_{LS} & Y_{LL} + R_L^{-1} \end{bmatrix} \quad (43)$$

Given that $1_{n+m}^T \Gamma_2 1_{n+m} = \sum_{i=n+1}^{n+m} r_i^{-1} < 0$, $\Gamma_2$ must have at least one negative eigenvalue. In fact, $R_L^{-1} = \Theta\Phi$ when $u_L = u_L^*$. According to (22), we obtain $u_L^* \succ \dfrac{u_{\text{ref}}}{2}1_m$, thus, $4u_{\text{ref}}^{-2}Y_1^{-1}\Theta \succ -Y_1^{-1}R_1^{-1} \succ O$. Then, according to Lemma 3 and (20), $\rho\left(-Y_1^{-1}R_1^{-1}\right) < 4u_{\text{ref}}^{-2}\chi < 1$.

Let $R_1 = I + Y_1^{-1}R_L^{-1}$, whose eigenvalues clearly have a positive real part. Thus, we have

$$R_2 = Y_1^{1/2}R_1Y_1^{-1/2} = I + Y_1^{-1/2}R_L^{-1}Y_1^{-1/2} > 0. \quad (44)$$

In fact, $Y_1 + R_L^{-1} = Y_1^{1/2}R_2Y_1^{1/2}$, and according to (44), $Y_1 + R_L^{-1} > 0$. Since $Y_{LS}\left(K^{-1} + Y_{SS}\right)^{-1}Y_{SL}$ is positive definite, according to eigenvalue perturbations theorem, we have

$$Y_{LL} + R_L^{-1} = Y_1 + R_L^{-1} + Y_{LS}\left(K^{-1} + Y_{SS}\right)^{-1}Y_{SL} > 0, \quad (45)$$

i.e., $Y_{LL}+R_L^{-1}$ is positive definite. Next, as $\Gamma_2$ has at least one negative eigenvalue and $Y_{LL}+R_L^{-1}$ is positive definite, according to Schur's theorem [23], $Y_{eq}$ must have at least one negative eigenvalue, thus completing the proof.

**Corollary 2.** For given $K$ and $u_{\text{ref}}$ that satisfy (26) and (27), the equilibrium is stable if

$$0<b<\min\left\{-\frac{C_{\min}}{\lambda_1(Y_{eq})},-\frac{1}{\lambda_1(Y_{eq})k_{\max}}\right\}, \quad (46)$$

where $C_{\min}=\min\{C_i\}$ and $k_{\max}=\max\{k_i\}$.

**Proof.** According to the eigenvalue perturbation theorem, (41) holds if the following is satisfied:

$$\begin{cases} C_{\min}+b\lambda_1(Y_{eq})>0 \\ k_{\max}^{-1}+b\lambda_1(Y_{eq})>0 \end{cases} \quad (47)$$

Thus, (46) is obtained, completing the proof.

Overall, the system admits a stable equilibrium when following these two steps:

**Step 1.** For given $P$ and $Y$, selecting the appropriate $u_{\text{ref}}$ and $K$ to ensure a constant steady state according to Theorem 2 and Corollary 1.

**Step 2.** Selecting the appropriate $b$ to ensure the equilibrium is stable according to Corollary 2.

## VI. CASE STUDY

To verify the presented analyses, we simulate a DC microgrid with the structure in Fig.1 using MATLAB/Simulink. The ideal CPL is modeled as a controlled current source, and the system parameters are listed in Table I.

### A. Stabilization Design

According to the proposed stabilization controller, the converter duty cycles are designed according to (38). Take $k_1 = k_2 = k_3 = k_4 = 1$, then, the duty cycles take the form

$$d_i = \frac{2}{3\times 10^5 b}\left(u_{\text{ref}}-\tilde{i}_{L_i}-u_i\right)+\frac{u_i}{300} \quad i\in\{1,2,3,4\}$$

where $b$ and $u_{\text{ref}}$ are the parameters that should be selected to guarantee that the system admits a stable equilibrium.

### B. Voltage Reference to Guarantee the Existence of Equilibrium

System equilibrium $u_L^*$ is determined by $U_L(\beta+Y_1 u_L)=-P$, which can be expressed as $u_L = F(u_L) = \zeta - Y_1^{-1}\Theta g(u_L)$, where $\zeta = u_{\text{ref}}[1\ 1\ 1\ 1\ 1\ 1]^T$. According to the resistances in Table I, $Y_1$ is given by

$$Y_1 = \begin{bmatrix} 1.5 & -1 & 0 & 0 & 0 & 0 \\ -1 & 11 & -5 & 0 & -5 & 0 \\ 0 & -5 & 7.5 & -2 & 0 & 0 \\ 0 & 0 & -2 & 2.833 & 0 & 0 \\ 0 & -5 & 0 & 0 & 7 & -2 \\ 0 & 0 & 0 & 0 & -2 & 2.667 \end{bmatrix}$$

According to Theorem 2 and Corollary 1, provided that either (26) or (27) holds, there must exist a unique vector $u_L^* \in D$ such that $u_L^* = F(u_L^*)$, i.e., the system admits a constant steady state. Let

$$\tau_1 = 2\sqrt{\chi},\ \tau_2 = \min_q\left\{\max_{1\leq i,j\leq m}\{f_{ij}(q)\}\right\},\ \tau_3 = \frac{\bar{\eta}+\underline{\eta}}{\sqrt{\bar{\eta}\underline{\eta}}}\sqrt{\chi},\ \tau_4 = 2\sqrt{\|Y_1^{-1}\Theta\|_\infty}$$

We use $q^*$ to denote the optimal vector that minimizes $\max_{1\leq i,j\leq m}\{f_{ij}(q)\}$. If $u_{\text{ref}} > \tau_2$, the system admits an equilibrium. Then, to test the correctness of the existence condition of euilibirium, we evaluated four cases:

**Case 1:** $P=10^3[1\ 1\ 1\ 0.5\ 0.5\ 0.5]^T$, $\tau_1=89.28$, $\tau_2=89.63$, $\tau_3=90.6$, $\tau_4=92.19$, $q^* = [2.3130\ 2.3167\ 2.1302\ 1.7741\ 2.2724\ 1.9308]^T$, $u_{\text{ref}}=89.64$;

**Case 2:** $P=10^3[1\ 1\ 1\ 0.5\ 0.5\ 0.5]^T$, $u_{\text{ref}}=89.6$;

**Case 3:** $P=10^3[2\ 2\ 2\ 1.5\ 1.5\ 1.5]^T$, $\tau_1=134.93$, $\tau_2=135.50$, $\tau_3=136.67$, $\tau_4=140.39$, $q^*=[1.4284\ 1.519\ 1.396\ 1.1887\ 1.532\ 1.3268]$, $u_{\text{ref}}=135.51$;

**Case 4:** $P=10^3[2\ 2\ 2\ 1.5\ 1.5\ 1.5]^T$, $u_{\text{ref}}=135.4$.

The corresponding results, obtained from MATLAB, are listed in Table II.

TABLE I.
Simulated System Parameters

| Parameter | Symbol(unit) | Value |
|---|---|---|
| Line resistance | $r_{ij}$ ($\Omega$) | $r_{15}=r_{56}=r_{37}=1$, $r_{67}=r_{28}=r_{69}=0.2$, $r_{78}=r_{9,10}=r_{4,10}=0.5$. |
| Converter input voltage | $V_i$(V) | $V_1=V_2=V_3=V_4=300$ |
| Converter inductances and capacitance | $L_i$ (mH), $C_i$ (mF) | $L_1=L_2=L_3=L_4=2$, $C_1=C_2=C_3=2$, $C_4=2.5$. |

TABLE II.
Equilibrium of the Evaluated Cases

| Cases | The corresponding $h\xi$, $\zeta$ such that $h\xi \prec F(h\xi) \prec F(\zeta) \prec \zeta$ | The solution of equation $u_L=F(u_L)$ |
|---|---|---|
| Case 1 | $\zeta=89.64[1\ 1\ 1\ 1\ 1\ 1]^T$, $h\xi=[43.25\ 43.18\ 46.96\ 56.39\ 44.03\ 51.82]^T$ | $u_L=[43.57\ 43.49\ 47.24\ 56.59\ 44.33\ 52.05]^T$ |
| Case 2 | Inexistence | Unsolvable |
| Case 3 | $\zeta=135.51[1\ 1\ 1\ 1\ 1\ 1]^T$; $h\xi=[70.03\ 65.85\ 71.65\ 84.15\ 65.29\ 75.39]^T$ | $u_L=[70.16\ 65.99\ 71.77\ 84.23\ 65.43\ 75.50]^T$ |
| Case 4 | Inexistence | Unsolvable |

The results shows that equation $u_L=F(u_L)$ has a solution if $u_{\text{ref}} > \tau_2$, and it may have no solution otherwise. In the evaluated cases, $\tau_2 < \tau_3 < \tau_4$, which shows that solvability condition (26) and (27) are stronger than the result in [22]. Moreover, the results shows that there is a solution when $u_{\text{ref}} >135.51$ and no solution when $u_{\text{ref}} =135.4$. Hence, condition (26) is less conservative.

### C. Performances of the proposed stabilization method

According to Corollary 2, the equilibrium is stable if (46) holds. Let us define $b_0$ as

$$b_0 = \min\left\{-\frac{C_{\min}}{\lambda_1(Y_{eq})},-\frac{1}{\lambda_1(Y_{eq})k_{\max}}\right\}$$

Hence, the system equilibrium is stable if $b<b_0$. To verify the correctness of the existence and stability condition of euilibirium, we evaluated three cases:

**Case 5.** $P=10^3[1\ 1\ 1\ 0.5\ 0.5\ 0.5]^T$, $u_{\text{ref}} =200$, $b_0 =0.062$, $b=3\times10^{-3}$.

**Case 6.** $P =10^3[0.5\ 0.5\ 0.5\ 0.5\ 0.5\ 0.5]^T$ for $t < 0.05$ and $P = 10^3[1\ 1\ 1\ 0.5\ 0.5\ 0.5]^T$ for $t \geq 0.05$, $u_{\text{ref}}= 89.64$, $b_0 = 2.15 \times 10^{-3}$, $b=1\times10^{-3}$;

**Case 7:** $P$ as in case 6, $u_{\text{ref}} = 89.6$, $b_0 = 2.15\times10^{-3}$, $b = 3\times10^{-3}$;

Moreover, CPLs were activated at $t=0.001$s in all these cases. In Case 5, $k_i=0$ for $t <0.2$ and $k_i=1$ onwards, i.e., the proposed stabilization acted for $t \geq 0.2$. In Case 6 and 7, $k_i=1$ throughout the simulation. The simulation results are depicted in Fig.3–5.

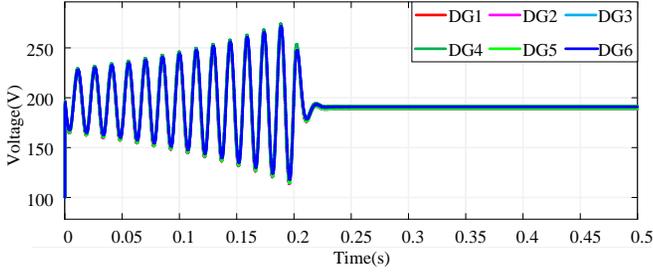
Fig.3. Load voltages for Case 5

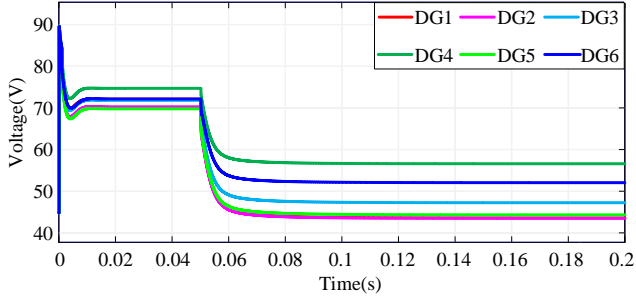
Fig.4. Load voltages for Case 6, $u_{ref} > \tau_2$

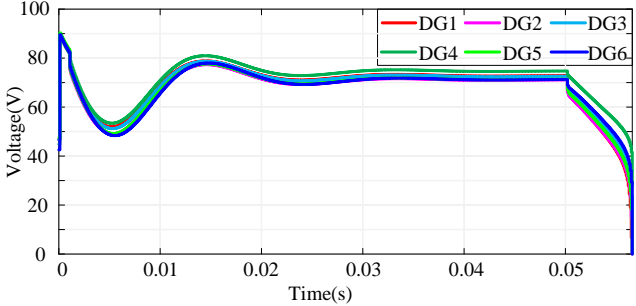
Fig.5. Load voltages for Case 7, $u_{ref} < \tau_2$

In Fig. 3, the system is unstable for $t < 0.2$ and stabilize after activating the proposed stabilization method, verifying its effectiveness. In Case 6, $u_{ref} > \tau_2$ and the system admits a stable equilibrium when the loads increase to maximal values as shown in Fig.4. In contrast, in Case 7, $u_{ref} < \tau_2$ and the load voltages collapse as shown in Fig.5. These results confirm the correctness of the sufficient conditions to the existence and stability of equilibrium presented in this paper.

## VII. Conclusions

We investigate the existence and stability of equilibrium of in general DC microgrids with multiple CPLs. A stabilization method is proposed and the sufficient conditions for the system admitting a stable equilibrium are derived. We transform the problem of nonlinear equation solvability into the existence of fixed-point of an increasing mapping and obtain the sufficient condition based on Tarski fixed-point theorem. The sufficient condition is less conservative comparing with the existing results. We adopt the linearized equivalent model around the equilibrium and obtained the stability conditions by analyzing the eigenvalue of the Jacobian matrix. These conditions provide a design guideline to build reliable DC microgrids. Finally, the simulation results verify the correctness of the proposed conditions.

## Appendix. Proof of Theorem 2.

**Proof.** According to 2) in Proposition 3, $Y_1^{-1}$ is strictly positive, and hence $Y_1^{-1}\Theta$ is also strictly positive. Consequently, $F(x_1) - F(x_2) = Y_1^{-1}\Theta(g(x_2) - g(x_1)) \succ 0_m$ for every $x_1 \succ x_2 \succ 0_m$, satisfying 1) of Lemma 2. Likewise, the system admits an equilibrium if 2) of Lemma 2 is also satisfied. Let $x_2 = \zeta$ and $x_1 = h\xi$, where $\xi = [q_1^{-1} \; q_2^{-1} \; \cdots \; q_m^{-1}]^T$ and $h$ is an undetermined positive scalar. Given that $Y_1^{-1}\Theta$ is positive, the following can be obtained:

$$F(\zeta) \prec \zeta . \quad (48)$$

Then, the quadratic equation in (14) is solvable if

$$h\xi \prec F(h\xi) . \quad (49)$$

Clearly, (49) can be expressed as

$$\frac{h}{q_i} < u_{ref} - \frac{1}{h}a_i k, \; i = 1, 2, \cdots, m, \quad (50)$$

and (50) is equivalent to

$$\begin{cases} \frac{q_1}{2}\left(u_{ref} - \sqrt{u_{ref}^2 - 4\frac{a_1 q}{q_1}}\right) < h < \frac{q_1}{2}\left(u_{ref} + \sqrt{u_{ref}^2 - 4\frac{a_1 q}{q_1}}\right) \\ \quad\quad\quad\quad \vdots \\ \frac{q_m}{2}\left(u_{ref} - \sqrt{u_{ref}^2 - 4\frac{a_m q}{q_m}}\right) < h < \frac{q_m}{2}\left(u_{ref} + \sqrt{u_{ref}^2 - 4\frac{a_m q}{q_m}}\right) \end{cases} . \quad (51)$$

Next, let

$$\Omega = \bigcap_{i=1}^{m}\Omega_i, \Omega_i = \left(\frac{q_i}{2}\left(u_{ref} - \sqrt{u_{ref}^2 - 4\frac{a_i q}{q_i}}\right), \frac{q_i}{2}\left(u_{ref} + \sqrt{u_{ref}^2 - 4\frac{a_i q}{q_i}}\right)\right).$$

If $\Omega \neq \varnothing$ (i.e., $F(\zeta) < \zeta$ and $F(h\xi) > h\xi$), according to Lemma 2, there must exist a unique vector, $h\xi \prec u_L^* \prec \zeta$, such that $u_L^* = F(u_L^*)$.

Clearly, $\Omega$ is non-empty if and only if

$$\begin{cases} \frac{q_i}{2}\left(u_{ref} - \sqrt{u_{ref}^2 - 4\frac{a_i q}{q_i}}\right) < \frac{q_j}{2}\left(u_{ref} + \sqrt{u_{ref}^2 - 4\frac{a_j q}{q_j}}\right) \\ \frac{q_j}{2}\left(u_{ref} - \sqrt{u_{ref}^2 - 4\frac{a_j q}{q_j}}\right) < \frac{q_i}{2}\left(u_{ref} + \sqrt{u_{ref}^2 - 4\frac{a_i q}{q_i}}\right) \end{cases} \quad (52)$$

holds for every $i, j \in \{1, 2, \ldots, m\}$. For specific $i$ and $j$, if $q_i = q_j$, (52) is solvable as

$$u_{ref}^2 > 4\max\left\{\frac{a_i q}{q_i}, \frac{a_j q}{q_j}\right\} \quad (53)$$

If $q_i \neq q_j$, (52) can be expressed as

$$2\left(\frac{a_i q}{q_j} + \frac{a_j q}{q_i}\right) - u_{ref}^2 < \sqrt{u_{ref}^2 - 4\frac{a_i q}{q_i}}\sqrt{u_{ref}^2 - 4\frac{a_j q}{q_j}} \quad (54)$$

By solving (54), the solution of (52) is given by

$$\begin{cases} u_{ref}^2 > 4\max\left\{\frac{a_i q}{q_i}, \frac{a_j q}{q_j}\right\} \quad \frac{a_i q}{q_j} + \frac{a_j q}{q_i} \leq 2\max\left\{\frac{a_i q}{q_i}, \frac{a_j q}{q_j}\right\} \\ u_{ref}^2 > \frac{\left(\frac{a_i q}{q_j} - \frac{a_j q}{q_i}\right)^2}{\left(\frac{a_i q}{q_j} + \frac{a_j q}{q_i} - \frac{a_i q}{q_i} - \frac{a_j q}{q_j}\right)} \quad \frac{a_i q}{q_j} + \frac{a_j q}{q_i} > 2\max\left\{\frac{a_i q}{q_i}, \frac{a_j q}{q_j}\right\} \end{cases}, \quad (55)$$

thus, obtaining (25) and completing the proof.